# THE NEW LIMIT FOR THE MAGNETIC MOMENT OF THE ELECTRON ANTINEUTRINO ($\bar{v}_e$) FROM THE MUNU EXPERIMENT

Z. DARAKTCHIEVA for MUNU collaboration
*Institut de Physique, A.-L.Breguet 1,
CH-2000 Neuchâtel, Switzerland*

The MUNU experiment uses the electron antineutrino-electron ($\bar{v}_e - e$) elastic scattering to examine the magnetic properties of the electron antineutrino ($\bar{v}_e$). The detector is situated at 18 m from the reactor, serving as an electron antineutrino source. Data corresponding to 66.6 days reactor-on and 16.7 days reactor-off are reported in this study. The electron recoil spectrum is obtained, using the visual scanning procedure. A new limit on the neutrino magnetic moment is derived: $\mu_v < 1.0 \times 10^{-10} \mu_B$.

## 1   Theoretical motivations

The Standard Model postulates for the neutrinos tiny magnetic moments, depending on the masses. There exist models [1], however, which predict magnetic moments of order of $10^{-10}\mu_B \div 10^{-12}\mu_B$. Neutrinos with such moments would undergo a spin flavour precession (SFP) in the magnetic field of the sun [2,3], complicating the pattern resulting from flavour oscillations [4,5]. To probe in particular the neutrino magnetic moment, one can study the electron antineutrino-electron ($\bar{v}_e - e$) elastic scattering, the cross-section of which is given by:

$$\frac{d\sigma}{dT_e} = \frac{G_F^2 m_e}{2\pi} \cdot \left[ (g_V + g_A)^2 + (g_V - g_A)^2 \cdot \left(1 - \frac{T_e}{E_\nu}\right)^2 + (g_A^2 - g_V^2) \cdot \frac{m_e T_e}{E_\nu^2} \right] +$$

$$+ \frac{\pi \alpha^2 \mu_\nu^2}{m_e^2} \cdot \frac{1 - \frac{T_e}{E_\nu}}{T_e} \qquad g_V = 2\sin^2\theta_W + \frac{1}{2}$$

$$g_A = -\frac{1}{2}$$

( 1 )

where $E_\nu$ is the neutrino energy, $T_e$ is the electron recoil energy, $\mu_\nu$ is the measured magnetic moment, $g_A$ and $g_V$ are the coupling constants. In the above formula, the first line represents the weak contribution of the cross section (from the Standard model), while the second line is the contribution from the magnetic moment $\mu_\nu$ [6]. The relative contribution of the magnetic moment term increases with a decrease of the neutrino energy and the electron recoil energy.

## 2  MUNU experiment

The electron antineutrino ($\bar{\nu}_e$) source of the MUNU experiment [7,8,9] is the commercial nuclear reactor in Bugey (France), having a power of 2800 MWth and emitting neutrinos in the low energy region (0 ÷ 8 MeV). The detector is situated at 18 m from the core of the reactor and has an overburden of 20 m.w.e. The neutrino flux in the lab is $10^{13}$ ν.cm$^{-2}$.s$^{-1}$

Figure 1, shows the MUNU detector, consisting of tree parts: the Time Projection Chamber, the Anti Compton detector and the Passive shielding.

The first and the main part of the detector is the Time Projection Chamber (TPC), filled with CF4 gas at 3 bar pressure. This gas was chosen, because of his good drifting properties, high density (3.68 g/l at 1 bar) and relatively low Z. The gas is contained in a transparent cylindrical acrylic vessel with length of 160 cm and diameter of 90 cm. The TPC acts as a detector and a target at the same time.

The second part of the detector is the Anti-Compton detector, which surrounds the TPC. It consists of 8 m$^3$ liquid scintillator and serves to veto cosmic muons and Compton electrons. The scintillator is contained in a steel vessel and is viewed by 48 photomultipliers.

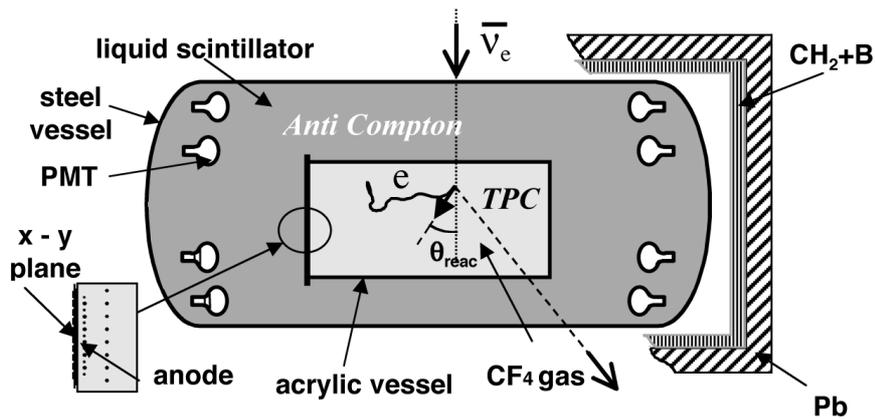

Figure 1: The MUNU detector.

The third part of the detector is the Passive shielding, serving to absorb the external neutrons and gammas. It has 15 cm Pb and 8 cm $CH_2$+B.

The tracking capability of the MUNU detector is demonstrated in Figure 2. The energy, angular and spatial resolutions of the TPC are found to be 8 % (1 $\sigma$) at 1 MeV, 10°, and 1.7 mm, respectively.

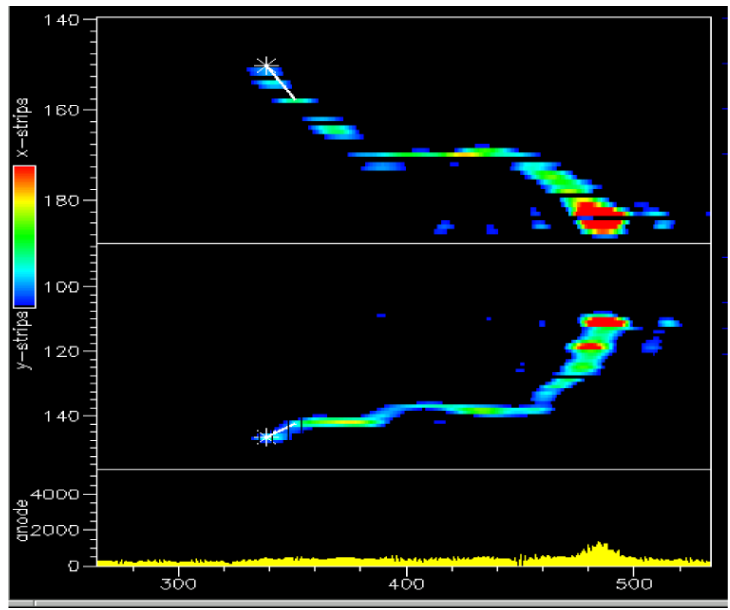

Figure 2 : A 670 kev electron in the TPC is shown. The fits of the first cm of the tracks on x-z and y-z projections are presented. The integrated anode signal is shown at the bottom of the picture.

## 3    Analysis and results

The analysis of the data is done at two steps. Firstly, the Compton electrons, muons, discharges, alphas, cosmics and uncontained electrons are rejected by automatic filtering. Secondly, the fully contained single electrons are selected.

The tangent of the first centimetres of the track is determined in a visual scanning procedure. From this fit one obtains the scattering angle with respect of the reactor axis $\theta_{reac}$, the angle with respect of the detector axis $\theta_{det}$ and the electron recoil energy $T_e$.

By using this method data of 66.6 days reactor-on and 16.7 days reactor-off were analysed. The following cuts have been applied: energy cut ($T_e >$ 700keV), angular cut ($\theta_{det} < 90^0$), kinematical cuts ($E_\nu > 0$, $\cos\theta_{reac} > 0$ for the forward electrons and $E_\nu > 0$, $\cos(\pi-\theta_{reac}) < 0$ for the backward electrons).

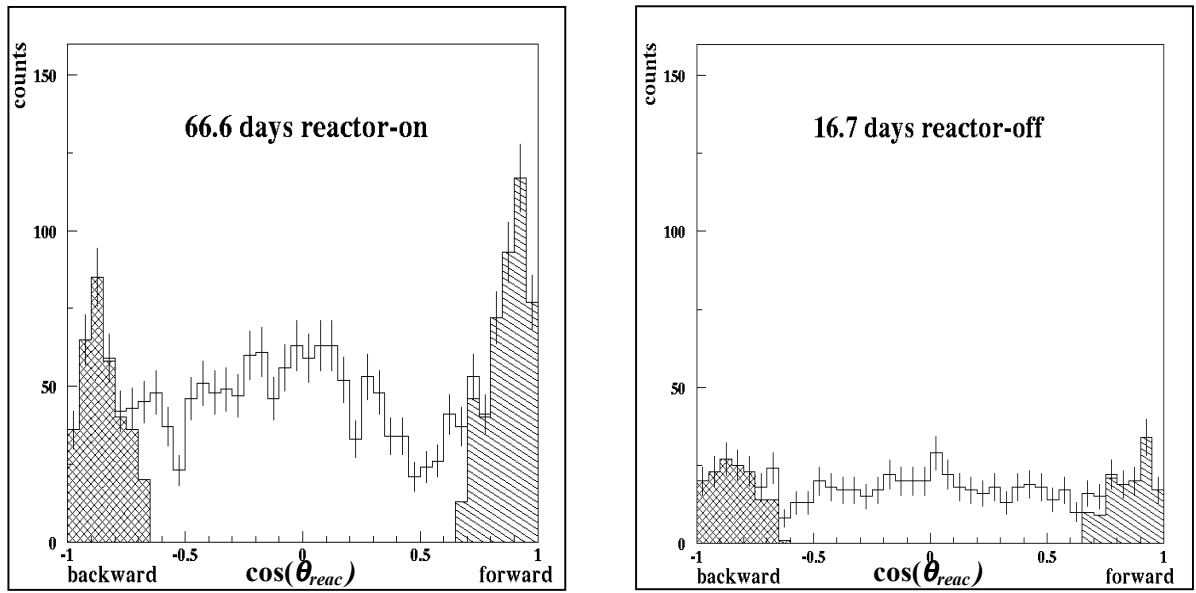

Figure 3: The distribution of $\cos\theta_{reac}$ for single electrons: reactor-on (left), reactor-off (right)

The angular distribution of $\cos\theta_{reac}$ of single electrons for the reactor-on and the reactor-off are shown on Figure 3. An excess of the events on the reactor side for the reactor-on period is seen: 458 forward events over 340 backward events. The cross checks with reactor-off gives: 130 forward electrons over 147 backward electrons. The integrated forward minus backward rate above 700 keV for the reactor-off is: -1.0 ±1.0 cpd, fully consistent with zero. The energy distribution of both the forward and the backward electrons for the reactor-on is shown on Figure.4. The energy distribution of forward minus backward rector-on spectrum is shown in Figure 5. The forward minus backward counts rate for energy higher than 700 keV is found to be 118 counts for 66.6 days, corresponding to 1.77±0.42cpd.

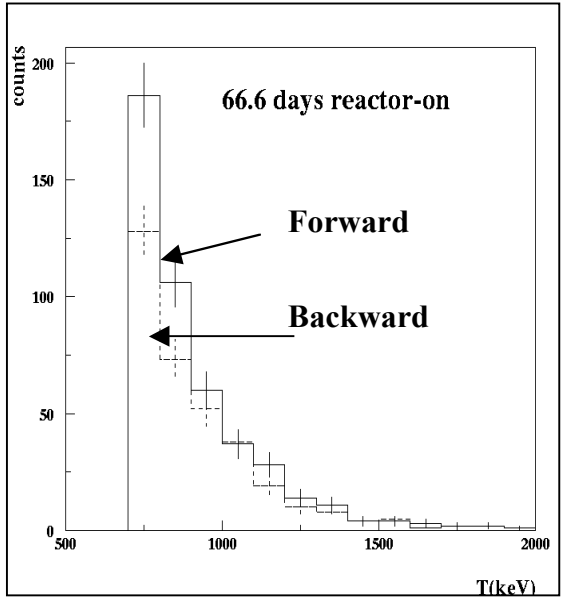 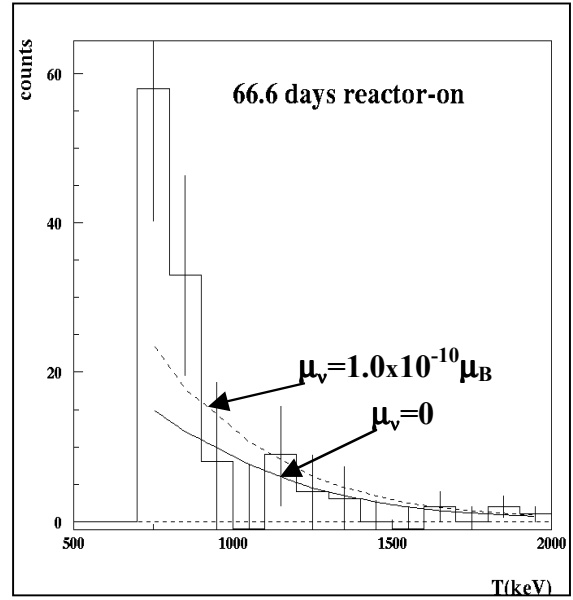

Figure 4: Energy distribution of forward and backward electrons reactor-on

Figure 5: Energy distribution of forward minus backward electrons reactor-on

A difference between the measured (1.77 ± 0.42 cpd) and expected (1.0 ± 0.1 cpd, with $\mu_v=0$) rates is observed for the electron energy higher than 700 keV.

The results in Figure 5 are difficult to explain with the neutrino magnetic moment, since the excess of the events is only at the first two channels: 700 keV and 800 keV. We note that, the neutrino spectrum in the energy region 0÷1.8 MeV is not as well known as in the higher energy range. It has been predicted from calculations only [10], and contributes significantly in the electron recoil spectrum below 900 keV.

For these reasons, we restricted our analysis to $T_e \geq 900$ keV, which corresponds essentially to the neutrino energy higher than 1.8 MeV. This part of the neutrino spectrum is known with a precision of the order of 5 % from direct measurements [11]. For $T_e \geq 900$ keV, we have a good agreement between the measured (0.41 ± 0.26 cpd) and expected (0.62 ± 0.05 cpd) rates.

The limit $\mu_\nu < 1.0 \times 10^{-10} \mu_B$ (90% C.L.) was obtained by using a simple statistical procedure, based on the chi square method.

This result improves on the existing limits. First that from the direct reactor measurement obtained in the Rovno experiment [12]: $\mu_\nu < 1.9 \times 10^{-10} \mu_B$. And then that from solar neutrinos, with the Super Kamiokande detector, yielding [13] the limit: $\mu_\nu < 1.5 \times 10^{-10} \mu_B$ (90% C.L.). Note that in Super Kamiokande experiment the solar neutrino was measured, but not the electron antineutrino.

## 4  Conclusion

The electron recoil spectrum was recorded by MUNU experiment.

The upper limit of the magnetic moment of the electron antineutrino is derived [14] : $\mu_\nu < 1.0 \times 10^{-10} \mu_B$ (90% C.L.).

### References

1. E. Akhmedov and J.Pulido, hep-ph/0209192,2002
2. M. Voloshin, M.Vysotski, L.Okun, Sov. Phys. JETP **64** (1986) 446.
3. E. Akhmedov et al.,A.Lanza,S.Petkov, Phys.Lett.B **348** (1995) 1658
4. B.Chanhan, J.Pulido, Phys.Rev D **66**(2002), 053006
5. B.Chanhan, J.Pulido, Torrente-Lujan, hep-ph/030497,2003
6. P.Vogel and J.Engel,Phys. Rev. D **39** (1989) 3378
7. The MUNU collaboration (C. Amsler et al.), Nucl. Inst. and Meth. A **396** (1997) 115
8 The MUNU collaboration (M. Avenier et al.), Nucl. Inst. and Meth. A **482** (2002) 408
9. The MUNU collaboration (C. Amsler et al.), Phys.Lett.B **545** (2002) 57
10. V. Kopeikin et al., Phys. of Atomic Nuclei Vol. **60** (1997) 172
11. G. Zacek et al., Phys. Rev. D **34** (1986) 2621
12. A. Derbin et al., JETP Lett. **57** (1993) 768
13. J. Beacom and P. Vogel, Phys. Rev. Lett. **83** (1999) 5222
14. The MUNU collaboration (Z.Daraktchieva et al), hep-ex/0304011, 2003.Accepted for publication on Phys.Lett.B